\documentclass[12pt,preprint]{aastex}

\usepackage{verbatim}
\usepackage{xspace}

\def\approx{$\sim$}
\def\persqcm{$\rm cm^{-2}$}

\def\h2{$\rm H_2$}
\def\error{$\pm$\xspace}
\def\e#1{$\times 10^{#1}$}
\def\tenup#1{10$^{#1}$}
\def\asec{\arcsec}

\def\deg{\arcdeg}

\def\kms{km~s$^{-1}$\xspace}

\def\solmass{$\rm M_{\sun}$\xspace}

\newcommand{\jykms}{$\rm Jy\,km\,s^{-1}$}
\newcommand{\mjb}{$\rm mJy\,b^{-1}$}
\newcommand{\halpha}{H$\alpha$\xspace}
\newcommand{\fhthree}{$f_{h3}$\xspace}
\newcommand{\fhfour}{$f_{h4}$\xspace}
\newcommand{\kmspasec}{$\rm km\;s^{-1}\;asec^{-1}$}

\begin{document}

\title{Star Formation and the ISM in Four Dwarf Irregular Galaxies}
\author{L. M. Young}
\affil{Physics Department, New Mexico Institute of Mining and Technology,
Socorro, NM 87801}
\email{lyoung@physics.nmt.edu}
\author{L. van Zee}
\affil{Astronomy Department, Indiana University, Bloomington, IN 47405}
\email{vanzee@astro.indiana.edu}
\author{K. Y. Lo}
\affil{Institute of Astronomy and Astrophysics, Academia Sinica, Taipei, Taiwan
ROC; and National Radio Astronomy Observatory, Charlottesville, VA 22903}
\email{flo@nrao.edu}
\author{R. C. Dohm-Palmer}
\affil{Astronomy Department, University of Minnesota, Minneapolis, MN 55455}
\email{rdpalmer@astro.umn.edu}
\author{Michelle E. Beierle}
\affil{Physics Department, New Mexico Institute of Mining and Technology,
Socorro, NM 87801}
\email{mbeierle@nmt.edu}

\begin{abstract}  
We present new, high sensitivity VLA observations of HI in four dwarf galaxies
(UGCA 292, GR8, DDO 210, and DDO 216)
and we use these data to study interactions between star formation and the
interstellar medium.
HI velocity dispersions and line shapes in UGCA 292, GR8, and DDO 210 show
evidence that 
these three galaxies contain both warm and cool or cold HI phases.
The presence of the cold neutral medium is indicated by a low-dispersion (3--6
\kms) HI component or by the Gauss-Hermite shape parameter $h_4 > 0$.
Contrary to expectations, we find no trend between the incidence of the
low-dispersion (colder) phase and the star formation rate in five dwarf
galaxies.
The colder HI phase may be a necessary ingredient for star formation, but it is
clearly not sufficient.
However, there is a global trend between the star formation rate of a galaxy and
the incidence of asymmetric HI profiles.  This trend probably reflects kinetic
energy input from young massive stars.
Numerical simulations show that the effects of rotational broadening (finite angular
resolution) are minimal for these galaxies.
Simulations are also used to estimate the errors in the column densities of the
high-dispersion and the low-dispersion HI phases.
\end{abstract}

\keywords{
galaxies: individual (UGCA 292, GR 8, DDO 210, DDO 216) --- 
galaxies: kinematics and dynamics ---
galaxies: dwarf
}

\section{Introduction} 

Among the dwarf galaxies we find an impressive variety of 
optical morphologies, colors, gas contents, 
star formation rates and gas depletion timescales.
\citet{youngblood1999} found a variation of more than three orders of
magnitude in the star formation rate per unit area of nearby dwarf irregulars.
The star formation rates of individual galaxies have varied by factors of more
than 10, both increasing and decreasing, over a Hubble time \citep{mateo1998}.
The Local Group contains some galaxies with evidence for recent star formation
but no sign of neutral gas, such as the Fornax dwarf spheroidal
\citep{stetson1998, young1999}.
It also contains some relatively gas-rich galaxies with little to no known
star formation activity, such as DDO 210 and DDO 216 \citep{vanzee2000}.
What happens in the interstellar medium (ISM) to cause such wide
variations in the gas consumption timescales?
The interplay between gas properties and star formation is one of the
more interesting unsolved issues in the evolution of dwarf galaxies.
We investigate this issue by careful study of the properties of the ISM in a
sample of four dwarf irregular galaxies and comparison with 
their current star formation activity.

Star formation activity in dwarf galaxies is expected to take place in molecular
gas, just as it does in our own Galaxy.
However, molecular gas is normally not directly detectable in the lowest mass (lowest
metallicity) dwarf galaxies \citep{taylor1998, vidal2000}. 
On the other hand, studies of our own Galaxy show that one can learn a great deal
about the ISM from HI.
\citet{clark1965} and \citet{radhakrishnan1972} found that
Galactic HI emission profiles often have a high-dispersion (10 \kms) optically
thin HI component superposed on a low-dispersion (3 \kms) optically thick component.
The low-dispersion component also appears in absorption against a continuum
source, whereas the high-dispersion component does not commonly appear in absorption.
The spatial distribution of the low-dispersion component is highly clumped;
it does not appear along every line of sight, but the high dispersion component
is ubiquitous.

Emission and absorption spectra can be used to estimate the spin temperatures of
these components.  The low dispersion component arises in a cold neutral medium 
(CNM) of
temperature \approx 100 K, and the high dispersion component arises in a
warm neutral medium (WNM) of temperature few thousand K \citep{kulkarni1988}.  A
recent review of observational and theoretical evidence is given by
\citet{wolfire2002}.
There is some evidence that a portion of the WNM may be in a thermally unstable
temperature range $ 500 \lesssim T_{kin} \lesssim 5000$ K, rather than the thermally
stable temperature range $ 5000 \lesssim T_{kin} \lesssim 8000$  K \citep{heiles2002,
kritsuk2002, gazol2001}.
However, there seems to be no disagreement that Galactic HI encompasses a
variety of temperatures.

The key connection between this HI work and what we hope to learn about star
formation in the dwarf galaxies is that {\it the 
cold phase of HI is usually assumed to be a necessary ingredient for star
formation} 
\citep{wolfire2002, parravano1988, parravano1989,
dickey2000, elmegreen2002a}. 
The reason is that molecular gas must form out of the cold HI rather than 
the warm \citep{glover2002}.
Indeed, \citet{elmegreen2002b} asserts that star formation must inevitably occur if a
cold phase is present.
In our own Galaxy there is evidence that the CNM is intimately
associated with molecular gas, as the abundance of the CNM peaks
in the molecular ring \citep{dickey2002}.
Thus, the amount and distribution of the CNM in dwarf galaxies may reveal
something about their star formation potential.

The CNM/WNM phase structure has been studied in many extragalactic systems via 
the traditional emission/absorption techniques 
\citep{dickey1993, braun1992, mebold1997, dickey2000}.
The two phases can also be detected in HI emission spectra by virtue of the
fact that they have different velocity dispersions; even when they occur at
nearly the same radial velocity, the superposition produces a distinct
non-Gaussian line shape with a narrow peak and broad wings.
In this way the cold and warm HI phases have been detected in a number of 
nearby spirals \citep{braun1997} and 500 pc below the plane of our own Galaxy 
\citep{lockman2002}.
\citet{braun1997} shows the spatial distribution of the HI associated
with the CNM in nearby spirals.
Many high velocity clouds also show evidence for CNM and WNM; clumpy HI with
linewidths $\lesssim 5$ \kms\ is superposed on more smoothly distributed HI with 
larger linewidths 
\citep{wakker1991, deheij2002, braun2000}.

Detailed studies of HI line profiles in the nearby dwarf galaxies Leo A and Sag
DIG \citep{young1996, young1997} (YL96, YL97) show features remarkably similar to 
Galactic HI profiles.
A broad HI component with a dispersion $\sigma$ \approx\ 8--10 \kms\ is
ubiquitous in these galaxies.  A narrower HI component of $\sigma$
\approx\ 3--5 \kms\ is concentrated into cloud-like structures of size
200--300 
pc, located usually (but not always) near regions of star formation 
activity in Leo A and Sag DIG.
Thus, the velocity dispersions and the spatial distributions of the two HI
components are exactly analogous to the properties of Galactic WNM and CNM.
\citet{sternberg2002} have constructed theoretical models of the ISM in Leo
A and Sag DIG, assuming that the gas must be in hydrostatic equilibrium in the
galaxies' gravitational potentials and assuming heating and cooling rates appropriate for
their metallicities.
These theoretical models predict the coexistence of cold and warm HI phases in
the dwarfs, in agreement with our observations.
Sternberg et al.\ have also made novel use of the HI properties of Leo A and 
Sag DIG to constrain the shape of the dwarfs' dark matter halos.

The evidence described above suggests that the ISM in nearby dwarf galaxies
does indeed contain CNM and WNM phases of HI in accordance with theoretical
models.
Furthermore, the two phases can be distinguished in HI emission spectra.
The CNM is not the raw material for star formation itself
but it is commonly assumed to be the raw material from which molecular gas is
made.
Therefore, we hypothesize that dwarf galaxies which contain greater amounts of CNM
should have higher star formation rates, in the same way and for the same
reasons that the molecular gas contents of
spiral galaxies are positively correlated with their \halpha or far-IR
luminosities \citep{kennicutt1998}.
The primary focus of this paper is to test that hypothesis.

We investigate the properties of the HI gas in four nearby dwarf
irregular galaxies through careful study of their HI line profiles.
This paper presents the highest quality HI images (best sensitivity and spectral
resolution) of DDO 216 (Pegasus dIrr), DDO 210, UGCA 292 (CVn
dwarf A), and GR8 (DDO 155).  
The HI line shapes in three of these dwarfs are parametrized by fitting
the profiles with two Gaussian components of different dispersions and also
with Gauss-Hermite polynomials.
We study the effects of ISM properties on star formation, and the effects
of star formation on ISM properties, by comparing the HI line shapes to star
formation activity on local (spatially resolved) and global (galaxy vs. galaxy)
scales.
We also present numerical simulations which quantify the reliability of the
double-Gaussian decomposition and the effects of rotational broadening.

\section{Sample Selection}

Using the line profiles to reveal the properties of the interstellar medium
requires, among other things, high signal-to-noise spectra (Appendix \ref{sntesting})
which are negligibly distorted by beam smearing effects.  Beam smearing
distorts the shape of the line profile when there is a large velocity gradient
within one angular resolution element \citep{takamiya2002, gentile2002}.
Thus, we selected nearby HI-bright galaxies with projected rotation velocities
comparable to or smaller than 20 \kms, the intrinsic width of the HI line in
dwarf galaxies \citep{lo1993, hoffman1996, vanzee1997a}.

In order to gain perspective on the relationship between star formation and the
ISM, and to test the hypothesis that more of the CNM
would be found in galaxies with higher star formation rates,
we also selected galaxies to span a range of \halpha\ luminosities.
The resulting sample includes two galaxies of high star formation rate
(GR8, UGCA 292) and two galaxies of low star formation rate (DDO 210, DDO
216).
Basic properties of these galaxies are found in Table \ref{sampletable}.
DDO 216 and DDO 210 are sometimes classified 
as ``transition" systems with optical properties intermediate between those of
dwarf irregulars and dwarf
spheroidals \citep{mateo1998}.  Other members of this class include the Antlia
dwarf, Phoenix, and LGS 3.
UGCA 292 is the most distant member of the sample, and the only one which
is not considered a member of the Local Group.
Analyses of the recent star formation histories of GR8 and DDO 216 have
been carried out by \citet{dp1998} and \citet{gallagher1998}, respectively.
\citet{tolstoy2000} show a new, deep color-magnitude diagram of DDO 210.
\citet{dp2003}
discuss the recent star formation history of UGCA
292.

\section{Observations and Data Reduction}

\subsection{HI data} 

All four galaxies were observed with the 
National Radio Astronomy Observatory's Very Large Array (VLA)\footnote{The 
National Radio Astronomy Observatory
is operated by Associated Universities, Inc., under cooperative agreement with
the National Science Foundation.} in its C and D configurations in 1995 and
1999.
These new data offer greatly improved sensitivity over older observations
\citep{lo1993, carignan1990} which were made in the mid-1980s.
Table \ref{obstable} gives specific dates, configurations, and times on source
for our new data.
Each galaxy was observed in one pointing centered roughly on the optical
center of the galaxy.  
Nearby point sources were observed every 30 to 45 minutes for use as phase calibrators.
The absolute flux scale was set by observations of the sources 0137+331 or 1331+305
(whichever was closer to the galaxy in question) and bandpass calibration was 
determined from those same sources.
All data calibration and image formation was done using standard calibration
tasks in the AIPS package.
Initial imaging revealed which channel ranges were free of HI line emission.
Continuum emission was subtracted directly from the raw uv-data by making first
order fits to the line-free channels.

The calibrated data were Fourier transformed using several different uvdata weighting
schemes chosen to enhance the spatial resolution or the sensitivity to large-scale
structures.  Dirty images were cleaned down to a residual level of 
0.8 to 1.0 times the rms noise fluctuations.
Table \ref{maptable} shows the velocity range covered for each galaxy as well
as the linear resolution (FWHM of the synthesized beam) and rms noise
level in the final image cubes.
All galaxies were observed at 1.3 \kms\ velocity resolution.

To determine if the total flux density was recovered in the HI synthesis
observations, total integrated flux profiles were constructed from the 
lowest resolution HI data cubes using the GIPSY task {\small \rm FLUX}.
Figure \ref{intspectra} shows integrated HI spectra for all four galaxies.
Their integrated HI fluxes, center velocities, and line widths are given in
Table \ref{HIfluxtable}.
In all four cases our integrated HI fluxes are greater than published single
dish fluxes.
Our HI flux for GR8 is within 10\% of that measured by \citet{tift1988}, and
the difference can be plausibly attributed to noise and/or calibration
uncertainties.
The single dish HI flux of UGCA 292 was measured by \citet{vanzee1997b} using
the wrong coordinates.
The velocities of DDO 210 and DDO 216 are so close to zero that confusion with
Galactic HI renders single dish HI fluxes highly uncertain.
Thus, there is no convincing evidence that our VLA maps have missed very
extended distributions.
The four sample galaxies span the full range of M(HI)/L$_B$ that are known for
dwarf galaxies, from the gas-rich UGCA 292 with M(HI)/L$_B$ = 7 to the gas-poor
DDO 216 with M(HI)/L$_B$ = 0.3.

Moment maps of the data cubes are shown in Figures \ref{cvnaolays} through
\ref{ddo216olays}.  The moment maps
were constructed from blanked cubes, where the signal was identified based
on spatial continuity between channels.  In order to include even the faint,
low signal--to--noise emission, each data cube was smoothed to a
resolution
of twice the beam prior to automatic clipping at the 2$\sigma$ level.
The resultant cubes were then interactively
blanked to remove spurious noise spikes.  A conditional transfer was applied
to
blank the corresponding locations in the original data cubes.  Moment
maps of the blanked cubes were created with the GIPSY task {\small \rm
MOMENTS}.

\subsection{Optical Broadband and H$\alpha$ data} \label{hasection}

Optical images of the four galaxies in this sample were obtained during
several observing runs at KPNO.\footnote{Kitt Peak National Observatory
is part of the National Optical Astronomy Observatories that are
operated by the Association of Universities for Research in Astronomy, Inc.
under contract to the National Science Foundation.}
The images for three of the four galaxies in this sample are presented in
\citet{vanzee2000}; briefly, broad band and narrow band images of DDO 210, DDO
216,
and UGCA 292 were obtained with the KPNO 0.9m telescope with the T2KA CCD.
The images have a spatial scale of 0.688\arcsec~per pixel, and typical seeing
of 1.5--2.0\arcsec.  The optical images of GR 8 used in this paper, kindly
provided by J. J. Salzer, were also obtained with the T2KA CCD on the KPNO
0.9m
and have similar depth and spatial resolution.
Plate solutions for the optical images were derived from coordinates of at
least 10 stars listed in the APM
catalog\footnote{http://www.ast.cam.ac.uk/$\sim$apmcat/}
and are accurate to 0.5\arcsec.
	   
As in \citet{vanzee2001}, the H$\alpha$ luminosities listed in Table
\ref{sampletable} were
calculated
from the observed H$\alpha$ flux within large apertures 
and thus contain H$\alpha$
emission from both diffuse and concentrated HII regions.  On average, the
diffuse emission contributes approximately 50\% of the total H$\alpha$
emission
from dwarf irregular galaxies \citep{youngblood1999, vanzee2000}.
Star formation rates were calculated from the H$\alpha$ luminosities using
the conversion factor from \citet{kennicutt1998}:
\begin{equation}
{\rm SFR} = 7.9 \times 10^{-42}~{\rm L(H\alpha)~M_{\odot}~yr^{-1}}.
\end{equation}

The current star formation rates of the galaxies in this sample range from
near zero
(DDO 210, SFR $<$ 0.000003 M$_{\odot}$ yr$^{-1}$ and DDO 216, SFR $\sim$
0.00003 M$_{\odot}$ yr$^{-1}$)
to a few thousandths of a solar mass per year (UGCA 292, SFR $\sim$ 0.0023
M$_{\odot}$ yr$^{-1}$
and GR 8, SFR $\sim$ 0.0040 M$_{\odot}$ yr$^{-1}$).
In fact, the estimated star formation rates of DDO 210 and DDO 216 are so
low that statistical uncertainties in their star
formation rates are substantially larger than the measurement 
uncertainties in the \halpha\ fluxes of Table \ref{sampletable}.
In a standard Salpeter initial mass function, only 6\% of the total mass in young stars
(0.1 to 100 \solmass) occurs in stars of mass greater than 25 \solmass.
The star formation rates of DDO 210 and DDO 216 are small
enough that only one or less than one HII region might be expected to
be present at any time, with the corresponding problems of small number
statistics. 
\citet{gallagher1998} also find that the star formation rate averaged over
the last 100 Myr in DDO 216 may be a factor of a few higher than the current
rate derived from the \halpha\ luminosity.

\section{Results}

\subsection{HI distributions}

Figures \ref{cvnaolays} through \ref{ddo216olays} show HI distributions,
overlays on optical images, and velocity fields of the four galaxies.
As illustrated by these overlays, the new data cubes are significantly
deeper than the data published in \citet{lo1993}; in particular, the
new observations of DDO 210 and DDO 216 indicate HI extents larger than
those shown in \citet{lo1993}.  The HI image of GR 8
is similar to that presented in \citet{carignan1990}, although the
present data have higher velocity resolution and are slightly deeper.

At first glance, the HI distributions appear very similar to the neutral
gas distribution in other dwarf irregular galaxies \citep{broeils1994, broeils1997,
vanzee1997b, swaters2002}.
In each case, the HI extends well beyond the 
optical image (excluding possible faint stellar halos).  The
HI--to--optical
ratio (measured at the 10$^{20}$ atoms cm$^{-2}$ and 25 mag arcsec$^{-2}$
isophotes, respectively)
is 2.3 for GR 8, 2.5 for UGCA 292, 2.1 for DDO 210, and 1.6 for DDO 216.
The peak column densities are also typical of dIs, reaching 1.3 $\times$
10$^{21}$ atoms cm$^{-2}$
in GR 8 and DDO 210, 3.7 $\times$ 10$^{21}$ atoms cm$^{-2}$ in UGCA 292, and
only
9.2 $\times$ 10$^{20}$ atoms cm$^{-2}$ in DDO 216 in the HI maps with
approximately
200 pc spatial resolution.  The central column density in UGCA 292 is similar
to those
found in BCDs \citep{vanzee1998, vanzee2001}, while GR 8, DDO 210, and DDO
216
have peak column densities similar to typical dIs \citep{vanzee1997b}.

The high spatial resolution images (lower left panel of Figures \ref{cvnaolays} -
\ref{ddo216olays})
indicate a rough correspondence between column density peaks and sites of
active star formation, including the faint HII knot in DDO 210 (see \citet{vanzee1997a}
for a spectrum of this enigmatic HII region).  As discussed in more detail
below
for the cold neutral phase, a high column density appears to be a necessary,
but not sufficient,
condition for star formation activity.

\subsection{HI kinematics} \label{kinematics}

While the gas distributions in these four galaxies appear to be typical of
dIs, the
gas kinematics of DDO 216 and GR 8 are somewhat unusual.  All four of these
galaxies
are at the low mass end of the gas--rich dwarf irregular class, and thus one
expects
the rotational component to be comparable to the velocity dispersion.  In DDO
210
and UGCA 292, a clear velocity gradient is visible in both the data cubes and
in the
derived velocity fields (see lower right panel of Figures \ref{cvnaolays} and
\ref{ddo210olays}).
The inclination corrected amplitudes of these velocity gradients are small,
0.31 \kms per arcsec and 0.10 \kms per arcsec for UGCA 292 ($i$ = 45\arcdeg)
and
DDO 210 ($i$ = 60\arcdeg), respectively,
but are non-negligible.  Velocity gradients are also present
in the velocity fields of GR 8  and DDO 216 (Figures \ref{gr8olays} and
\ref{ddo216olays}).  However, the gas kinematics are
not as well
ordered in these two galaxies; rather than a smooth velocity gradient from
one side to
the other, DDO 216 and GR 8 appear to have clumps of gas with similar
kinematic
properties  (2 distinct clumps in DDO 216 and 3 clumps in GR 8).  In GR8, the three
clumps can be identified in the HI column density images, and in DDO 216 they can be
seen in a major axis position-velocity diagram (Figure \ref{ddo216pv}) at $-200$ \kms\
and at $-180$ \kms. The gas
kinematics
appear well ordered within each clump, and there is an overall kinematic
profile along
the optical major axis of both of these galaxies. Nonetheless, the overall
impression is of
random gas motions rather than a rotating disk in DDO 216 and GR 8.  

In these low mass
galaxies, the kinematic motions may be complicated by expanding shells and
bubbles,
which obscure the global dynamics.  For example, in the position-velocity diagram of
Figure \ref{ddo216pv} a circular structure centered at
$-190$ \kms\ and $-1.5'$ (1.5\arcmin\ northwest of the center) may be evidence for an
expanding HI bubble of radius \approx 1\arcmin\ (200 pc) and expansion velocity 10 \kms.
The dynamical age of such a structure would be 2\e{7} yr, but it is curious that there
is very little star formation activity of this vintage in the galaxy 
\citep{vanzee2000, gallagher1998}.

\subsection{Line shapes and velocity dispersions}\label{shaperesults}

Because the line profiles in DDO 216 are double-peaked throughout much of the
galaxy (Figure \ref{ddo216pv}), we do not attempt a detailed analysis of the velocity dispersion or
profile shapes in that galaxy.
But among the other three galaxies we find the usual variety of line shapes,
including some which are 
well described by single Gaussian components and others which are not.  
Example spectra are shown in Figures \ref{puregauss}, \ref{h4spec}, and
\ref{h3spec}.  
Deviations from pure Gaussian shapes take the form of broader
wings and a narrower peak than a simple Gaussian, and sometimes asymmetry,
just as has been found for other dwarf galaxies (YL96, YL97) 
as well as nearby spirals \citep{braun1997}.

All spectra in these galaxies were fit with a single Gaussian component.
Profiles were also fit with a superposition of two Gaussians of different
dispersions, as was done in YL96 and YL97. 
In addition, we fit the HI profiles with Gauss-Hermite polynomials,
which are commonly used for parametrizing the line-of-sight velocity
distributions of elliptical galaxies.
The Gauss-Hermite polynomials (up to fourth order) are
$$ \phi(x)=a\: e^{-y^2/2}\: \left\{ 1+ \frac{h_3}{\sqrt{6}}(2\sqrt{2}y^3 - 
    3\sqrt{2}y) + \frac{h_4}{\sqrt{24}}(4y^4 - 12y^2 + 3) \right\}, $$
where $y \equiv (x-b)/c$ \citep{vdMF1993}.
These functions are essentially Gaussian terms modified by an asymmetric 
bit whose amplitude is indicated by $h_3$ and a symmetric bit whose amplitude 
is $h_4$.
In the case that $h_3 = h_4 = 0,$ the function is a Gaussian of amplitude $a$, 
center $b$, and dispersion $c$.  If $h_3 \ne 0$ the profile is asymmetric, and 
if $h_4 \ne 0$ it has either a more pointed ($h_4 > 0$) or a more flat ($h_4 <
0$) top than a Gaussian.
This parametrization is a useful alternative to the double-Gaussian
decomposition because the $h_3$ and $h_4$
polynomials are orthogonal, so that errors in those parameters are
uncorrelated (\citet{vdMF1993}, Appendix \ref{sntesting}). 
In contrast, when two superposed Gaussian components are fit simultaneously,
their parameters are highly correlated.
The disadvantage of the Gauss-Hermite formulation is that there is no simple
physical interpretation of the magnitudes of $h_3$ and $h_4$, whereas two
superposed Gaussians can be naturally interpreted in terms of gas of different
velocity dispersions and/or radial velocities.
All line profile fits were made using the {\small \rm XGAUFIT} task in GIPSY.

The procedure used to distinguish profiles which are simple Gaussians from those
that are not is described in YL96 and YL97. 
Briefly, our null hypothesis is that a simple Gaussian function is 
an equally good description of the profile as two Gaussians
or the Gauss-Hermite functions with terms up to $h_4$.
The rms of the residuals is computed for all three fits over
the velocity range of the line emission (roughly the central 40 of 120 channels).
We divide the rms of the residuals to a single Gaussian fit by
the rms of the residuals to a double Gaussian fit or a Gauss-Hermite fit.
We reject the null hypothesis at the 90\% confidence level
in a one-tailed f-test 
when the ratio of the rms's is 1.27 and at the 95\% confidence level when
the ratio of the rms's is 1.36, for 30 degrees of freedom.  

\subsubsection{Gauss-Hermite analysis}\label{gausshermite}

Figures  \ref{gr8h3hiha}, \ref{gr8h4hiha}, \ref{cvnah3hiha}, \ref{cvnah4hiha},
\ref{ddo210h3hiha}, and \ref{ddo210h4hiha} indicate where the Gauss-Hermite fits
show significant deviations from Gaussian line profiles.
In DDO 210 (Figure \ref{ddo210h4hiha}), a large fraction of the profiles have positive $h_4$ values
indicating broader wings and a narrower peak than a simple Gaussian.  Very few
of the profiles in DDO 210 show asymmetries (nonzero $h_3$; Figure
\ref{ddo210h3hiha}).

The three prominent HI clouds of GR8 each contain young massive stars and
\halpha emission; the stellar content of these regions is described by
\citet{dp1998}.
The southwestern clump contains the highest concentration of massive
main sequence stars and is therefore interpreted as the youngest region of star
formation activity in the galaxy.
This HI clump shows asymmetric profiles with $h_4 > 0$ (Figures \ref{gr8h3hiha} and
\ref{gr8h4hiha}), but only in the 
southern half of the clump near the star formation activity.
The eastern HI clump has the highest concentration of blue
helium-burning stars, which are somewhat older and less massive than main
sequence stars of comparable luminosity.  
This clump shows remarkably few departures from Gaussian profile shape.
The northern HI clump has the smallest massive star 
content, and it is interpreted as the oldest of the three regions of recent star
formation activity.
Line profiles throughout this entire HI clump have high degrees of asymmetry
(the highest $h_3$ values measured in the entire sample) and $h_4 > 0$.

In UGCA 292 (Figure \ref{cvnah3hiha}) we also find a large incidence of 
asymmetric profiles, $h_3 > 0$ in the western part of the galaxy and $h_3 < 0$
in the southeast.
In between those two regions of asymmetric profiles, and coincident with the
most luminous HII region in the galaxy, we find profiles which are
indistinguishable from pure Gaussians (Figure \ref{puregauss}).  These simple
Gaussian profiles have higher dispersions, 10--11 \kms, than are found elsewhere
in the galaxy (7.5--9.5 \kms).
Some regions of $h_4 > 0$ are found, generally offset by one beamwidth (300 pc)
from HII regions (Figure \ref{cvnah4hiha}).

UGCA 292 is the galaxy in this sample with the largest rotational
velocity, and if beam smearing were important it would be expected to cause 
an antisymmetric pattern similar to the observed one with $h_3 > 0$ on one side 
of the galaxy and $h_3 < 0$ on the other.
Therefore, we conducted experiments to determine the magnitude of beam smearing
effects in this galaxy.
The simulations described in Appendix \ref{rotationmodeling} indicate that
values of $h_3$ and $h_4$ no larger than 0.01 would be expected from beam
smearing effects alone in this galaxy, whereas the observed $h_3$ and $h_4$ are 
up to 0.1.  
Furthermore, if the $h_3$ pattern in this galaxy were caused by beam smearing we
would expect the line between $h_3 < 0$ and $h_3 > 0$ to lie along the kinematic
major axis at position angle +60\deg\ (Figure \ref{cvnaolays}).
Instead, that line lies 50\deg\ away at position angle +110\deg.
Thus, we believe the measured $h_3$ and $h_4$ values (and subsequent
double-Gaussian decomposition) are indicative of local
conditions in the gas rather than beam smearing.

\subsubsection{Double-Gaussian decomposition}

Figures \ref{cvnahist}, \ref{gr8hist}, and \ref{ddo210hist} show the fitted
dispersions in the three galaxies.  For profiles which are adequately 
described by a single Gaussian, that best-fit dispersion is plotted; for
profiles which require two components at greater than 90\% confidence, the two
dispersions are plotted independently.
In the double-Gaussian profiles, the narrower components have dispersions
typically 3--6 \kms\ and the broader ones are typically 8--13 \kms.
As in the cases of Leo A and Sag DIG, we find the notable result that the profiles
which require only one Gaussian component have dispersions very similar to
the broader component of the double-Gaussian profiles.
Thus, we again find a component of HI of dispersion \approx 10 \kms {\it
everywhere} throughout these galaxies.
In certain locations we also find an additional HI component of dispersion 3--5
\kms.
We interpret these results to mean that the HI in these galaxies contains a
ubiquitous high-dispersion phase, which we identify with the WNM, and a 
clumpy low-dispersion phase, which we identify with the CNM.

Comparison of Figures \ref{cvnahist}, \ref{gr8hist}, and \ref{ddo210hist}
suggests that both the high-dispersion and low-dispersion phases in
DDO 210 have dispersions lower by 1--2 \kms\ than their counterpart phases in
the other two galaxies.

Figures \ref{gr8narrow}, \ref{cvnanarrow}, and \ref{ddo210narrow} show the
spatial distribution of the low dispersion component (CNM) with respect to the total
HI column density and \halpha\ emission.
From these figures it is apparent that the set of profiles in which we find the
low dispersion component is almost identical to the set of profiles with $h_4 > 0$.
Specifically, the distribution of the low-dispersion component in GR8 (Figure \ref{gr8narrow})
is qualitatively similar to the image of nonzero $h_4$ in Figure
\ref{gr8h4hiha},  Figure \ref{ddo210narrow} is similar to Figure
\ref{ddo210h4hiha}, and Figure \ref{cvnanarrow} is similar to Figure
\ref{cvnah4hiha}.
The two techniques for describing 
deviations from simple Gaussian shapes give very similar results on
where the CNM is located.

\section{Discussion}

\subsection{Line shapes and star formation: small scale correlations}

We find some correspondence between the presence of the CNM and star formation
activity.
Specifically, all regions of \halpha emission in these three galaxies
are found within about one beamwidth or less of profiles containing the low
dispersion component (or having $h_4 > 0$).
In the case of the eastern HI clump of
GR8 and its ringlike HII region, the number of such profiles is small but
nonzero.
In the case of UGCA 292 we find the largest spatial offsets, 300 pc, between
current star formation activity and evidence of the CNM.
But it is generally true that for these galaxies and for the previously studied
Leo~A and Sag~DIG (YL96, YL97), 
some evidence of CNM is always near to current star formation activity.

The reverse of the previous statement is most definitely not true, however.
There is not always evidence of star formation activity near to regions
containing the CNM.
The most dramatic example is DDO 210, where \halpha imaging revealed 
very minimal star formation activity
(Table \ref{sampletable}; \citet{vanzee2000}) and the color-magnitude
diagram also indicates a very weakly populated upper main sequence
\citep{tolstoy2000}.
But DDO 210 contains ample amounts of the CNM, as indicated in Figures
\ref{ddo210h4hiha} and \ref{ddo210narrow}.
Other examples include the eastern portion of GR8 near RA = 12 58 42.0 (Figures
\ref{gr8h4hiha} and \ref{gr8narrow}), the northwestern quadrant of Leo~A
(Figure 9 of YL96), and the western half of Sag DIG (Figure 9 of
YL97).

In UGCA 292, the regions of star formation activity are all near (within about
one beamwidth) of asymmetric HI profiles (Figure \ref{cvnah3hiha}).
In GR8, the regions of star formation activity are all spatially coincident
with asymmetric HI profiles, with the possible exception of the eastern HI
clump and its ring HII region (Figure \ref{gr8h3hiha}).
In these two galaxies we find a rather close correspondence between asymmetric
profiles and star formation activity.
Two regions near the outer edges of DDO 210 show asymmetric
profiles in places where there is no evidence of current star formation
activity (Figure \ref{ddo210h3hiha}).
However, in the aggregate,
the evidence for small scale correlations between asymmetric profiles
($h_3 \ne 0$) and star formation activity is at least as good as, and perhaps
better than, the evidence for small scale correlations between $h_4$ and star
formation.

We do not see strong evidence of a temporal
evolution in the state of the ISM. 
As noted in section \ref{gausshermite}, the three regions of star formation activity in GR8 have been
approximately dated by their massive stellar content \citep{dp1998},
and we do not see a one-to-one relationship between the stellar content
and either the presence of CNM or the presence of asymmetric profiles.
As stars are born and die there most certainly will be some evolution
of the state of the ISM, but detecting it may require a larger dataset or one
which probes different spatial and temporal scales.

\subsection{Line shapes and star formation: Global correlations}

We estimate, for each galaxy, the fraction of profiles that show deviations from pure
Gaussian shapes.  The parameters \fhthree and  \fhfour (Table \ref{h3vshatable})
are the number of profiles for which a nonzero $h_3$ or $h_4$ are 
detected at greater than $3\sigma$ confidence, divided by the number of 
profiles in the galaxy with signal-to-noise ratio greater than 20.
The normalization is an attempt to account for differences in the overall
signal-to-noise ratio of the cubes and the numbers of pixels per
independent beam.  
For UGCA 292 and DDO 210, virtually all of the profiles with nonzero $h_3$ or $h_4$ have
signal-to-noise ratios greater than 20, and \fhthree\ and \fhfour\ are less than one.
However, note that strong $h_3$ and $h_4$ signatures can be detected even when the
profile has a total signal-to-noise ratio less than 20 (Table \ref{monte-h3}), so that 
\fhthree and \fhfour are not constrained to be less than one.
This effect is responsible for the large \fhthree\ and \fhfour\ in GR8.

Our data show a notable trend between a galaxy's \halpha\ luminosity
and the fraction of the profiles which are asymmetric.
For this analysis we have also included Leo A and Sag DIG, which we fitted with
Gauss-Hermite polynomials in exactly the same manner as described above.
Table \ref{h3vshatable} indicates, for each galaxy, the number of spectra with
signal-to-noise ratios greater than 20, the fraction of spectra with significant
measurements of nonzero $h_3$ and $h_4$, and the \halpha luminosities.
The data are plotted in Figures \ref{h3vssfr} and \ref{h4vssfr}, where it is
clear that the galaxies with larger \halpha\ luminosities have a greater fraction
of their profiles asymmetric.
In contrast, there is little correspondence between the fraction
of the profiles that show evidence for CNM ($h_4 > 0$) and the \halpha\
luminosity.

A galaxy-wide relation between the \halpha luminosity and the fraction of
profiles that show asymmetries could naturally arise if the young massive stars
are injecting kinetic energy into the ISM.
A young massive star located exactly in the midplane of a homogeneous gas
disk might be expected to accelerate the surrounding gas in a symmetric manner,
but stars outside of the midplane of real galaxies with inhomogeneous 
interstellar media
will be more likely to accelerate the surrounding gas more strongly in some
directions than in others.
The result will be an asymmetric HI velocity profile.

The fits to the spectrum in Figure \ref{h3spec} show that one can decompose
these asymmetric profiles into components typically separated by a few \kms.
We can thus obtain an order-of-magnitude estimate of the kinetic
energy represented by the asymmetric profiles of GR8, for example. 
If something like one third of the HI in GR8 is accelerated to velocities of 5
\kms\ with respect to the rest of the gas in the galaxy, that would require
8\e{50} erg of kinetic energy.
This value is a small fraction of the total thermal and/or turbulent energy in
the HI gas, since most of the HI in the galaxy has a velocity dispersion
between 8 and 13 \kms\ (Figure \ref{gr8hist}).
Assuming the energy input from supernovae and/or stellar winds is converted
into kinetic energy of the ISM at an efficiency of 10\% \citep{lozinskaya1992}, we
require \tenup{52} erg of input energy, or a few canonical supernovae,
to explain the asymmetric HI profiles.
Given that we know massive star formation is now occurring in GR8, the energy
requirements are not unreasonable. 
The interpretation that the asymmetric HI profiles are the result of kinetic
energy input from young massive stars is plausible in this respect.

The small number of asymmetric profiles in DDO 210 can also be easily understood, given
the extremely low rate of massive star formation activity.
\citet{maclow1999, maclow2002}
estimates the timescale for dissipation of turbulent kinetic energy
in the ISM to be on the order of 10 Myr.
Thus, we expect the signatures of kinetic energy injection ($h_3 \ne 0$) to disappear
very shortly after the massive stars and their HII regions disappear.
Furthermore, we speculate that one might directly infer the size scales on which turbulent 
kinetic energy is injected into the ISM from the size scales of contiguous regions
with similar $h_3$ values.  In the present set of galaxies, those size scales are
usually comparable to or smaller than our HI beam sizes (100--300 pc, Table \ref{maptable}).

This interpretation that the asymmetric profiles trace energy injection from
young massive stars has relied on the use of \halpha\ to trace the star
formation rate, as usual.
Section \ref{hasection} mentions, however, that the star formation rate of
DDO 210 in particular is so low that the \halpha\ emission may not be
a particularly accurate indicator of the total star formation activity.
But to the extent that the most massive stars are the primary sources of
kinetic energy input into the ISM, and their lifetimes are comparable to the
10 Myr timescale for dissipation of turbulent kinetic energy as mentioned
above, the \halpha\ emission of these galaxies is an indicator of the
instantaneous energy injection rate.  These considerations justify the use of 
\halpha\ luminosities in Figure \ref{h3vssfr}.
In an analysis of CNM as a raw material for star formation activity (Figure
\ref{h4vssfr}), we would indeed be more interested in the {\it total} star
formation rate rather than just the massive star formation rate.  But the
qualitative result coming out of Figure \ref{h4vssfr}, a lack of correlation
between \halpha\ luminosity and CNM content, would not be noticeably
affected even if DDO 210 were shifted by a factor of 10 in the horizontal
direction relative to the other galaxies.

\subsection{Star formation and the CNM}

It has sometimes been suggested that the broad (but symmetric) wings of 
non-Gaussian profiles --- the $h_4 > 0$ shape --- are caused by star formation 
activity stirring the gas \citep{dickey1990}.
That doesn't appear to be the case in these dwarf galaxies.
For example, \citet{young1996, young1997} showed that the HI velocity dispersions in Leo A and
Sag DIG actually {\it decrease} in the regions of $h_4 > 0$, rather than
increase.  This result is contrary to what one would
expect if the broad wings were due to extra kinetic energy.
Furthermore, we now have the very dramatic example of the many profiles in DDO
210 with $h_4 > 0$ even though there is little or no current star formation activity.
In this respect the ISM of DDO 210 is reminiscent of the ISM in high velocity
clouds \citep{wakker1991, braun2000, deheij2002}.

Rather, the evidence we have presented above suggests that the asymmetric
profiles ($h_3 \ne 0$) are the result of star formation activity, whereas
the narrow Gaussian component and its corresponding signature $h_4 > 0$ 
are due to the presence of the CNM.  
(Technically speaking, the measured velocity dispersions of the narrow Gaussian
component do not require it to be at temperatures \approx 100 K. 
A dispersion of 5 \kms\ would, if it were entirely thermal, correspond to a
temperature of 3000 K and place the gas in the thermally unstable region of the
phase diagram.  We would then expect the formation of cold gas from the unstable
gas, and in addition there may be nonthermal gas motions within our beam.)
However, it is clear that the presence of CNM does not automatically lead to
star formation.
The presence of CNM seems to be a {\it necessary} condition for star formation--
all regions of star formation activity have some CNM close by-- but the CNM is
clearly not a {\it sufficient} condition for star formation.

\citet{lo1993} discussed the HI properties of nine dwarf irregular
and transition dwarf galaxies, paying particular attention to the issue of 
their gas contents and star formation rates.  They argued that it is difficult to
understand the low star formation rates (both current and averaged over a Hubble
time) in the faint dwarfs which have low degrees of rotational support.
The present paper takes this argument one step further.
One possible reason for the low star formation rates is that only the CNM in
these galaxies, not the WNM, should be considered possible raw material for star
formation.  In Leo A and Sag DIG (YL96, YL97) we found that only 20\% or so of
the HI is CNM.
But for DDO 210, the mystery of the low star formation rate remains.
As far as we can determine without actually detecting molecular gas, the
conditions in the ISM of this galaxy seem favorable for the formation of stars.
Perhaps there is a significant time delay between the formation of the CNM and
the formation of the molecular gas-- or whatever else is the remaining necessary
ingredient for star formation.

The quantitative relationships between star formation activity and the CNM
are also probably complicated by the fact that energy injection into the ISM
may eventually destroy the CNM.  Such negative feedback may be responsible 
for the fact that most of the CNM in Figures \ref{gr8narrow} through
\ref{ddo210narrow} tends not to be located directly on top of the HII
regions.  This kind of negative feedback may also contribute to a lack of
correlation between the global CNM content of the galaxy and the \halpha\
luminosity.  If the CNM is destroyed by energy injection it will
probably become WNM, and indeed in this respect it is significant that the
models of \citet{sternberg2002} actually require some star formation activity in
order to produce WNM in the interiors of dwarf galaxies.
In other words, the interstellar pressures in the interiors of the model galaxies
are high enough to drive all HI into the CNM phase in the absence of some energy
injection.

\subsection{Constraints on dark matter halos}

The theoretical analyses of \citet{sternberg2002} incorporate radiative transfer,
thermal, and ionization balance to model hydrogen gas in hydrostatic equilibrium in the
gravitational potential of a small dwarf galaxy.  
The model galaxy contains both stars and dark matter appropriate for the Local Group
dwarfs Leo~A and Sag DIG, and the gas is heated and photoionized by an external
radiation field.
Models with either constant density dark matter cores or highly underconcentrated NFW
halos produce WNM column densities and projected HI scale heights which agree with
the observed properties of Leo~A and Sag DIG.

UGCA 292, GR8, and DDO 210 
are broadly similar to Leo~A and Sag DIG, so that all the conclusions which
Sternberg et al.\ have made about their dark matter halos are echoed by this new 
set of galaxies.  For example, the constant density core model favored by
Sternberg et al.\ has a peak WNM
column density of 1.4\e{21} \persqcm; for DDO 210, GR8, and UGCA 292 we find peak WNM
column densities of 1.0\e{21}, 1.2\e{21}, and 3.4\e{21} \persqcm\ respectively.
The higher column densities in UGCA 292 may simply be projection effects.
The model has a $1/e$ scale height in the projected HI column density of 0.6 kpc;
we find HI scale heights of 500 pc for GR8, 600 pc for UGCA 292, and 240 pc for DDO 210.
(These HI scale heights are measured as the geometric mean of the major and minor axes at
the $1/e$ contour in the HI column density.)
Thus, our new HI observations of DDO 210, GR8, and UGCA 292 show that they also
require either halos with
constant density cores or underconcentrated cuspy halos.  
The constraints may be less strong for DDO 210 than for the others, since DDO 210 is
somewhat smaller in luminous mass and its HI scale heights are only half as large.
The implications of these ``soft cores" for the nature of dark matter are still
being discussed (e.g. \citet{donghia2003} and others).

\section{Summary}

We present the highest sensitivity images of HI emission from the dwarf
irregular galaxies UGCA 292 and GR8 and from the transition dwarfs DDO 210 and
DDO 216.
The HI of DDO 216 is clearly resolved into two partially overlapping 
gas clouds which are separated by about 20 \kms\ in velocity.
Many of the HI profiles in that galaxy are double-peaked.

We have analyzed the HI line profile widths and shapes in DDO 210, GR8, and UGCA
292 by fitting with Gaussians and Gauss-Hermite polynomials.
Our double-Gaussian decomposition indicates that a HI component of dispersion
9--12 \kms\ is present everywhere in these three galaxies, and in some locations
a second component of dispersion 3--5 \kms\ is also present.
We also map out the locations of asymmetric HI profiles, indicated by $h_3 \ne
0$ in the Gauss-Hermite fits.
The two fitting techniques give consistent answers: line profiles which require two
Gaussians, one broad and one narrow, are also well described by $h_4 > 0$. 
We interpret the presence of the narrow Gaussian component ($h_4 > 0$) as
evidence for the CNM.
Of the six dwarf galaxies which have been studied in this way, five show
evidence for the CNM.

We have estimated the effects of rotational broadening (finite
angular resolution) on the line profiles of these galaxies.  Even for the galaxy
with the largest velocity gradient (UGCA 292), the rotationally-induced
deviations from simple Gaussian line shapes are much smaller than the deviations
which are actually observed. 

If the CNM is indeed the site of molecular gas formation and a necessary 
ingredient for star formation, one might expect a correlation between the
quantity of CNM in a galaxy and its star formation rate.
We estimated the amount of CNM in a galaxy by noting the fraction of profiles
with $h_4 > 0$.  
There is no apparent relation between that fraction and the \halpha\ luminosity 
of five dwarfs spanning more than three orders of magnitude in $\rm L_{H\alpha}$.
The most striking cases are DDO 210, where we find a large incidence of CNM but
very little star formation, and UGCA 292, where we find a relatively small
incidence of CNM but a large star formation rate.
The CNM may indeed be a necessary ingredient for star formation but it is
clearly not sufficient by itself to ensure star formation.

There is a trend between the fraction of asymmetric profiles in
a galaxy and $\rm L_{H\alpha}$, in the sense that galaxies with greater star
formation rates also have a greater fraction of asymmetric profiles.
This result probably indicates that star formation activity is responsible for
stirring the HI, and the kinetic energy involved in accelerating the HI to
typical speeds of 5 \kms\ is reasonable for galaxies with these kinds of star
formation rates.
Small scale correlations between the locations of HII regions and the regions of
asymmetric profiles are less clear:  in some cases the asymmetric
profiles are coincident with the HII regions, and in other cases the asymmetric
profiles appear around the edges of the HII regions.

The three main star formation regions of GR8 have been approximately dated from
their young stellar contents.
We do not see a clear evolutionary sequence in the properties of the ISM in
these three star formation regions.
A detection of this kind of temporal evolution may require larger sample sizes
which probe different spatial and/or temporal scales than what we are able to do
in the present work.

\acknowledgments

Thanks to J.\ J. Salzer for providing the broadband and \halpha\ images of GR8.
We appreciate the insightful comments from the referee.
Support for proposal number 9044 was provided by NASA through a grant from the
Space Telescope Science Institute, which is operated by the Association of
Universities for Research in Astronomy, Incorporated, under NASA contract
NAS5-26555.
LMY and LvZ thank the Institute of Astronomy and Astrophysics, Academia Sinica,
Taiwan, for travel support and hospitality.
KYL thanks the Academia Sinica and the National Science Council, Taiwan, for
research support.

\appendix

\section{Line shape measurements at low signal-to-noise ratios}\label{sntesting}

A proper understanding of the uncertainties in the fitted spectral line
parameters is crucial to the interpretation of most of the results in this
paper.
Therefore, we undertook a series of Monte Carlo simulations which were designed
to show (1) the reliability of the spectral line shape parameters
for profiles of low signal-to-noise ratio, and (2) the
degree of correlation between the fitted parameters.

We first constructed data cubes consisting of identical model spectra with
1.3 \kms\ velocity resolution.  The line
profiles were prescribed by the Gauss-Hermite formula given in Section
\ref{shaperesults} with $c = 6.2$ \kms, $h_3 = 0.05$, and $h_4 = 0.10$,
values typical for the observed galaxies.
Random, normally distributed noise was added to the simulated spectra in
order to achieve a specified signal-to-noise ratio.
For this work, signal-to-noise ratios are defined as the peak intensity of the line
divided by the rms noise level.
In this manner we constructed 16800 spectra at each of the signal-to-noise
ratios specified in Table \ref{monte-h3}.
The procedure was repeated for sets of profiles with the same signal-to-noise
ratios, $c$, and $h_4$ but with $h_3 = 0.002$.
All profiles were fit with the Gauss-Hermite polynomials.

Table \ref{monte-h3} gives the fitted values of $h_3$, $h_4$, and their standard
deviations as a function of signal-to-noise ratio.
For signal-to-noise ratios of 7 and higher, the fitted $h_3$ and $h_4$ are
normally distributed with standard deviations equal to the formal uncertainty
reported by the fitting routine.
At signal-to-noise ratios less than 7, the fitted $h_3$ and $h_4$ are no longer
normally distributed.
Therefore we conclude that for signal-to-noise ratios $\ge 7$, the formal
uncertainties in the fitted $h_3$ and $h_4$ are reliable estimates of their
errors.
We also find very convincing evidence that the $h_3$ and $h_4$ parameters are 
indeed independent of each other.
Specifically, we find no correlation between the fitted $h_3$ and the fitted
$h_4$ of a given profile.  The fitted $h_3$ values are also the same, within
their errors, regardless of whether or not the $h_4$ term is included in the
fit.
Finally, the fitted $h_4$ and the standard deviations in $h_3$ and $h_4$ 
were identical for the cases $h_3 = 0.05$ and $h_3 = 0.002$. 

Simulations of the double-Gaussian fitting procedure were made in a similar way.
We constructed data cubes of 4225 spectra each and 1.3 \kms\ velocity
resolution; each spectrum was filled with a superposition of two Gaussian
components of 4 \kms\ and 11 \kms\ dispersions.  The amplitude ratio of the two
components was fixed at 1.4:1 (narrow:broad).
Random noise was added to the spectra to achieve total signal-to-noise ratios 
of 30, 20, and 11, which is the range of signal-to-noise ratios occupied by the
non-Gaussian profiles in GR8.
In one set of data cubes the two components have the same velocity, and in a
parallel set of data cubes they have velocities offset by 5 \kms.
The spectra were then fit with a superposition of two Gaussian components.

The fitted amplitudes, center velocities, and dispersions of the two Gaussian
components along with their standard deviations are presented in Table
\ref{monte-gauss}.  
As for the Gauss-Hermite fits, we find that the standard deviations in the
fitted parameters are equal to the formal errors quoted by our fitting routine.
We conclude that the formal errors are indeed reliable measures of the
uncertainties in the parameters (at least at the signal-to-noise ratios studied
here).
We also note that at signal-to-noise ratios of 11, a 5 \kms\ velocity offset 
between the two components is not detectable at a 3$\sigma$ confidence level.
However, at signal-to-noise ratios of 20 and above this velocity offset is
easily measured.

Finally, we find that the fitted amplitudes of the two Gaussian components
are {\it strongly} correlated in the sense $ A_1 + A_2 =$ constant.
The standard deviations in the fitted amplitudes are primarily determined by
this correlation.
As a result, the standard deviations of the fitted amplitudes {\it do not}
decrease as the signal-to-noise ratio increases; rather, they are fixed at about
1.4 times the rms noise level.  The situation is mildly improved for asymmetric
profiles, in which cases the standard deviations in the fitted amplitudes are in
the range 1.0 to 1.2 times the rms noise.
The last column of Table \ref{monte-gauss} shows that the standard deviations in
the column densities of the two Gaussian components range from 10\% to about
40\%.
We would expect uncertainties in the column densities to be larger in cases
where the amplitude ratio is more extreme.

\section{Rotational broadening}\label{rotationmodeling}

When an object with a velocity gradient is observed at finite angular
resolution, its spectral line profiles appear artificially broadened.
The line profiles can also be asymmetric, which is responsible for the fact that
different rotation curves for galaxies are obtained from the peak of the
spectra, the median, and the mean velocity 
\citep{takamiya2002, gentile2002}.
Therefore, we conducted simple numerical experiments to determine whether the
values of $h_3$ and $h_4$ which we observe in the HI of dwarf galaxies could
be artifacts of their velocity gradients.

In our present sample the largest velocity gradient is in UGCA 292, where we
find 0.2 \kmspasec\ or 2.9 \kms per beam in the high
resolution cube and 3.5 \kms per beam in the low resolution cube.
The GIPSY task GALMOD was used to simulate observations of galaxies with 
various velocity fields and gas distributions, including the following:
\begin{itemize}
\item velocity gradients ranging from  0.3 to 1.2 \kmspasec
\item velocity fields which are solid body throughout and ones which rise
linearly to a radius of about 60\asec\ and flatten
\item uniform gas distributions as well as those that drop off by about a factor
of 20 between the center and the outside edge at 100\asec
\item intrinsic velocity dispersions between 0.1 \kms and 10 \kms.
In the absence of beam smearing, all HI spectra would be simple Gaussians of
this dispersion.
\end{itemize}
All model galaxies were ``observed" at 1.3 \kms velocity resolution and were
smoothed to 14\asec\ linear resolution, similar to the high resolution cube of
UGCA 292.

The synthetic spectra were fit with the Gauss-Hermite polynomials, which gave the
following results.
Models which closely approximate the parameters for UGCA 292 (0.3 \kmspasec\
velocity gradient, intrinsic velocity dispersion \approx 10 \kms) showed no
significant $h_3$ or $h_4$.  All of the fitted $h_3$ and $h_4$ values were less
than 0.01 and most were consistent with zero, within their uncertainties.
In comparison, in the real galaxies the well-measured $h_3$ and $h_4$ are between
0.05 and 0.25. 
Synthetic galaxies did not have artificial $h_3$ and $h_4$ values as large as
the ones found in the real galaxies unless the ratio of the velocity gradient to
the intrinsic velocity dispersion was more than 10 times larger than what is
found in the real galaxies.
We conclude that the beam smearing which is present in the real galaxies is not
large enough to explain the observed departures from simple Gaussian profiles.
Rather, these departures must be indicative of local conditions in the ISM.

The velocity gradients which are present in the real galaxies also make very minor
contributions to the observed line widths.
The fitted dispersions of the synthetic galaxies are roughly consistent with the
expectation
$ (\Delta v_{obs})^2 \sim (\Delta v_{true})^2 + (\Delta v_{grad})^2, $
where $\Delta v_{obs}$ is the observed line profile width, $\Delta v_{true}$ is
the intrinsic line width, and $\Delta v_{grad}$ is the product of the velocity
gradient and the beam size \citep{braun1997}. 
In UGCA 292, the velocity gradient amounts to about 3 \kms\
per beam; the observed line widths (FWHM) are \approx 12 \kms\ for the narrow
component and 23 \kms\ for the broad component.
Thus, the beam smearing of the velocity gradient contributes at most a few 
percent to the observed line widths.
In this paper we therefore consider that significant changes in the line shape
or width from one place to another must be indicative of changing conditions in
the ISM, not global effects connected with the velocity gradient.

\clearpage
\begin{deluxetable}{llcrcccc}
\tablewidth{0pt}
\tablecaption{Sample Galaxies
\label{sampletable}}
\tablehead{
\colhead{Galaxy} & \colhead{Alt. Name} & \colhead{RA} & \colhead{Dec} & 
\colhead{Distance}  & \colhead{M$_B$} & \colhead{L(\halpha)} &
\colhead{References} \\
\colhead{}       & \colhead{} & \multicolumn{2}{c}{J2000.0} & 
\colhead{Mpc} & \colhead{} & \colhead{$10^{36}$ erg s$^{-1}$} &
\colhead{}
}
\startdata
UGCA 292 & CVn dw A & 12 38 40.0 & 32 46 00 & 3.5 & $-11.67$ & 295 (3) &
     1, 2, 3 \\
GR 8  & DDO 155 & 12 58 39.4 & 14 13 02 & 2.2 & $-12.12$ & 500 (25) &
     4, 4, 5\\
DDO 210 & Aquarius dI & 20 46 57.3 & $-$12 49 50 & 0.95 & $-10.95$ & 0.37 (0.32)
&   6, 2, 3 \\
DDO 216 & Pegasus dI & 23 28 34.3 & 14 44 50 & 0.76 & $-12.20$ & 3.7 (0.4) &
    7, 2, 3 \\
\enddata
\tablecomments{The coordinates given here are the pointing centers used for the 
HI observations;
they roughly correspond to the optical centers of the galaxies.
All coordinates given in this paper are epoch J2000.0.
All \halpha luminosities are from reddening-corrected fluxes measured in
large (galaxy-size) apertures, so they include diffuse \halpha emission as well
as HII regions.
Uncertainties in the \halpha luminosities do not include distance;
the uncertainty in the \halpha\ flux of GR8 is not given by Youngblood \& Hunter
(1999) and is arbitrarily assumed to be 5\%.
The M$_B$ for UGCA 292 which is given in van Zee (2001) and the \halpha
luminosity of GR8 which is given in Youngblood \& Hunter (1999) have been
corrected to the distances assumed here.
References-
1) \citet{dp2003}; 
2) \citet{vanzee2001}; 
3) \citet{vanzee2000}; 
4) \citet{dp1998}; 
5) \citet{youngblood1999}; 
6) \citet{lee1999}; 
7) \citet{gallagher1998} 
}
\end{deluxetable}

\begin{deluxetable}{lccc}
\tablewidth{0pt}
\tablecaption{VLA HI Observations
\label{obstable}}
\tablehead{
\colhead{Galaxy} & \colhead{Config} & \colhead{Date} & 
    \colhead{Time on source} \\
\colhead{}       & \colhead{} & \colhead{} & \colhead{hr}
}
\startdata
UGCA 292 & Cs & 24Jan99 & 13.3 \\
GR 8  & Cs & 25Jan99 & 12.8 \\
DDO 210 & DnC & 20Jan95 & 2.1 \\
        & Cs  & 25Jan99 & 7.5 \\
DDO 216 & D & 13Mar95 & 2.2 \\
        & Cs & 24Jan99 & 12.7 \\
\enddata
\end{deluxetable}

\begin{deluxetable}{lcccc}
\tablewidth{0pt}
\tablecaption{HI maps
\label{maptable}}
\tablehead{
\colhead{Galaxy} & \colhead{Velocity Range} & 
    \multicolumn{2}{c}{Resolution} & \colhead{Noise level} \\
\colhead{} & \colhead{\kms} &  \colhead{\asec} &
    \colhead{pc} & \colhead{mJy}
}
\startdata
UGCA 292 & (383,231)  & 17.7 $\times$ 17.4 & 300 $\times$ 295 & 0.80 \\
         &               & 14.2 $\times$ 13.9 & 241 $\times$ 236 & 0.90 \\
GR 8  & (285, 145)  & 18.6 $\times$ 18.1 & 199 $\times$ 193 & 0.85 \\
      &             & 14.8 $\times$ 14.6 & 158 $\times$ 156 & 0.96 \\
DDO 210 & ($-66$, $-209$)  & 27.4 $\times$ 20.8 & 126 $\times$ 96 & 1.3 \\
        &                  & 20.8 $\times$ 16.0 & 96 $\times$ 74 & 1.5 \\
        &                  & 47.5 $\times$ 37.0 & 218 $\times$ 170 & 1.6 \\
DDO 216 & ($-103$, $-256$)  & 25.1 $\times$ 21.2 & 93 $\times$ 78 & 0.80 \\
        &                   & 17.3 $\times$ 15.9 & 64 $\times$ 59 & 0.90 \\
        &                   & 56.2 $\times$ 52.2 & 207 $\times$ 192 & 1.0  \\
\enddata
\end{deluxetable}

\begin{deluxetable}{lcccccc}
\tablewidth{0pt}
\tablecaption{Global HI parameters
\label{HIfluxtable}}
\tablehead{
\colhead{Galaxy} & \colhead{V$_{sys}$} & \colhead{$\Delta V_{50}$} &
    \colhead{$\Delta V_{20}$} & \colhead{HI Flux} & \colhead{M(HI)} &
    \colhead{M(HI)/L$_B$} \\
\colhead{} & \colhead{\kms} & \colhead{\kms} & \colhead{\kms} &
    \colhead{\jykms} & \colhead{\solmass} & \colhead{}
}
\startdata
UGCA 292 & 308.8 & 29.4 & 44.9 & 17.60 & 5.09\e{7} & 7.0 \\
GR 8  & 213.9 & 27.8 & 43.6 & 9.74 & 1.11\e{7} & 1.0 \\
DDO 210 & $-140.7$ & 21.8 & 35.5 & 15.18 & 3.23\e{6} & 0.87 \\
DDO 216 & $-183.3$ & 24.6 & 40.5 & 29.90 & 4.06\e{6} & 0.34 \\
\enddata
\end{deluxetable}

\begin{deluxetable}{lcccc}
\tablecolumns{5}
\tablewidth{0pt}
\tablecaption{
\label{h3vshatable}}
\tablehead{
\colhead{Galaxy} & \colhead{N$_{>20}$} & \colhead{\fhthree} & 
\colhead{\fhfour} & \colhead{L(\halpha)} \\
\colhead{} & \colhead{} & \colhead{} & \colhead{} & \colhead{$10^{36}$ erg s$^{-1}$}
}
\startdata
UGCA 292 & 250 (14.5) & 0.40 (0.05) & 0.31 (0.04) & 300 (3)\\
GR8     & 78  (4.1) & 0.96 (0.16) & 1.1 (0.17) & 500 (25)\\
DDO 210 & 251 (7.9) & 0.08 (0.02) & 0.77 (0.07) & $<$ 0.4 (0.3) \\
Leo A & 1037 (27.7) & 0.38 (0.02) & 0.93 (0.04) & 9.7 (0.5) \\
Sag DIG & 647 (15.5) & 0.28 (0.02) & 0.75 (0.05) & 8.1 ($^{+8.1}_{-4.1}$) \\
\enddata
\tablecomments{
N$_{>20}$ is the number of profiles with signal-to-noise ratio
(fitted profile maximum / rms in line-free regions) greater than 20.  
In parentheses is the number of beam areas corresponding to that number
of pixels.
The values in parentheses after \fhthree\ and \fhfour\ are estimates of their
uncertainties assuming Poisson statistics in counting the relevant numbers of
profiles.
The \halpha\ flux of Sag DIG is taken from \citet{strobel1991},
assuming a distance of 1.1 Mpc \citep{longmore1978, cook1988}, and
multiplied by a factor of two to account for a typical contribution of diffuse
\halpha\ emission \citep{vanzee2000}.  Consequently the uncertainty in its \halpha\
flux is a factor of two.
The \halpha\ flux of Leo A is taken from \citet{youngblood1999}, assuming a
5\% uncertainty and a distance of 0.69 Mpc \citep{tolstoy1998}. 
}
\end{deluxetable}

\begin{deluxetable}{ccc}
\tablecolumns{3}
\tablewidth{0pt}
\tablecaption{Recovery of $h_3$ and $h_4$ at low S/N
\label{monte-h3}}
\tablehead{
\colhead{Total S/N} & \colhead{Fitted h3} & 
\colhead{Fitted h4}  
}
\startdata

30 & 0.0501 \error 0.0124 &  0.0999 \error 0.0124  \\
20 & 0.0501 \error 0.0186 &  0.0998 \error 0.0186  \\
10 & 0.0504 \error 0.0378 &  0.0997 \error 0.0384  \\
7  & 0.0506 \error 0.0552 &  0.100 \error 0.0572  \\
5  & 0.0500 \error 0.0875 &  0.105 \error 0.0946  \\
\enddata
\end{deluxetable}

\begin{deluxetable}{cccrrc}
\tablecolumns{6}
\tablewidth{0pt}
\tablecaption{Double-Gaussian Decomposition {\it vs.} Line
Strength
\label{monte-gauss}}
\tablehead{
\colhead{Total S/N} & \colhead{Input Amp} & \colhead{Fitted Amp} &
\colhead{Center} & \colhead{Dispersion} & \colhead{Column Density} \\
\colhead{} & \colhead{} & \colhead{} & \colhead{\kms} & \colhead{\kms}
& \colhead{\% Error}
}
\startdata

\sidehead{Symmetric Profiles} 
30 & 17.65 & 17.65\error 1.41 & 0.00\error 0.16 & 3.99\error 0.28 & 14 \\
   & 12.32 & 12.32\error 1.41 & $-0.02$\error 0.40 & 11.04\error 0.67 & 13\\
   \\
20 & 11.76 & 11.76\error 1.41 & 0.00\error 0.25 & 3.98\error 0.43 & 21 \\
   & 8.24 & 8.26\error 1.46 & 0.00\error 0.63 & 11.10\error 1.00 & 19 \\
   \\
11 & 6.47 & 6.59\error 1.29 & 0.00\error 0.55 & 3.93\error 0.81 & 37 \\
   & 4.53 & 4.53\error 1.41 & $-$0.05\error 1.60 & 11.44\error 2.44 & 32 \\
   \\

\sidehead{Asymmetric Profiles}
30 & 17.65 & 17.72\error 1.05 & 0.00\error 0.18 & 4.00\error 0.24 & 11 \\
   & 12.35 & 12.35\error 0.98 & $-$0.10\error 0.60 & 11.00\error 0.44 & 9 \\
   \\
20 & 11.76 & 11.87\error 1.06 & 0.02\error 0.26 & 3.99\error 0.37 & 16 \\
   & 8.24 & 8.24\error 0.99 & $-$0.14\error 0.95 & 11.01\error 0.68 & 13 \\
   \\
11 & 6.47 & 6.67\error 1.18 & $-$0.10\error 1.50 & 3.98\error 0.71 & 32 \\
   & 4.53 & 4.55\error 1.12 & $-$0.50\error 2.30 & 10.96\error 1.51 & 24 \\
\enddata
\tablecomments{All amplitudes are given in units of the noise level in the spectra.
The top row in each pair gives values for the narrow component (input $\sigma =
4$ \kms) 
and the bottom row gives values for the broad component (11 \kms).
Asymmetric profiles had the broad component offset in velocity by 5 \kms\ from the
narrow component, and that offset has been subtracted from the fitted center velocities.
}
\end{deluxetable}

\clearpage
\begin{figure} 
\figcaption{Integrated HI spectra derived from the low resolution data cubes
for each galaxy.  These galaxies were selected for this project because
their global profiles have velocity widths at the 50\% level less than 30 km
s$^{-1}$.
\label{intspectra}
}
\end{figure}

\begin{figure} 
\figcaption{Moment maps of UGCA 292. (Top left) HI column density (contours) from the
lowest
resolution data cube overlayed on an image from the Digitized Sky Survey.  The
contours correspond to 0.3, 0.6, 1.2, 2.4, 4.8, and 9.6
$\times$ 10$^{20}$
atoms cm$^{-2}$. (Top right) Same, but with the column density shown in the
grey
scale as well as the contours. (Lower left) HI column density (contours) from the highest
resolution data cube overlayed on an H$\alpha$ image.  The contours
correspond to 1., 2., 4., 8., 16., and 32. $\times$ 10$^{20}$ atoms cm$^{-2}$.
(Lower right) Velocity field of the lowest resolution data cube.  The contours are marked
every 5 km s$^{-1}$.
\label{cvnaolays}
}
\end{figure}

\begin{figure} 
\figcaption{Moment maps of GR 8.  (Top left) HI colum density (contours) from the lowest
resolution data cube overlayed on an R-band image.  The
contours correspond to 0.3, 0.6, 1.2, 2.4, 4.8, and 9.6
$\times$ 10$^{20}$
atoms cm$^{-2}$. (Top right) Same, but with the column density shown in the
grey
scale as well as the contours. (Lower left) HI column density (contours) from the highest
resolution data cube overlayed on an H$\alpha$ image.  The contours
correspond to 1., 2., 4., and 8. $\times$ 10$^{20}$ atoms cm$^{-2}$ (note that
the inner most
contour corresponds to a {\it depression} in the HI distribution).
(Lower right) Velocity field of the lowest resolution data cube.  The contours are marked
every 5 km s$^{-1}$.
\label{gr8olays}
}
\end{figure}

\begin{figure} 
\figcaption{Moment maps of DDO 210.  (Top left) HI column density (contours) from the
lowest
resolution data cube overlayed on a B-band image.  The
contours correspond to 0.1, 0.2, 0.4, 0.8, 1.6, 3.2, 6.4, and
12.8 $\times$ 10$^{20}$
atoms cm$^{-2}$. (Top right) HI contours and grey scale for the medium resolution data
cube.
Contours correspond to 0.5, 1.0, 2.0, 4.0, and 8.0 $\times$
10$^{20}$ atoms cm$^{-2}$.
(Lower left) HI column density (contours) from the highest
resolution data cube overlayed on an H$\alpha$ image.  The contours
correspond to 1., 2., 4., 8., and 16. $\times$ 10$^{20}$ atoms cm$^{-2}$.
(Lower right) Velocity field of the medium resolution data cube.  The contours are marked
every 2.5 km s$^{-1}$.
\label{ddo210olays}
}
\end{figure}

\begin{figure} 
\figcaption{Moment maps of DDO 216.  (Top left) HI column density (contours) from the
lowest resolution data cube overlayed on a B-band image.  The
contours correspond to 0.1, 0.2, 0.4, 0.8, 1.6, 3.2, and 6.4
$\times$ 10$^{20}$
atoms cm$^{-2}$. (Top right) HI column density contours and grey scale for the medium resolution data
cube.
Contours correspond to 0.5, 1.0, 2.0, 4.0, and 8.0 $\times$
10$^{20}$ atoms cm$^{-2}$.
(Lower left) HI column density (contours) from the highest
resolution data cube overlayed on an H$\alpha$ image.  The contours
correspond to 1., 2., 4., and 8. $\times$ 10$^{20}$ atoms cm$^{-2}$.
(Lower right) Velocity field of the medium resolution data cube.  The contours are marked
every 5 km s$^{-1}$.
\label{ddo216olays}
}
\end{figure}

\begin{figure} 
 \includegraphics[scale=1.0,bb=72 42 535 343]{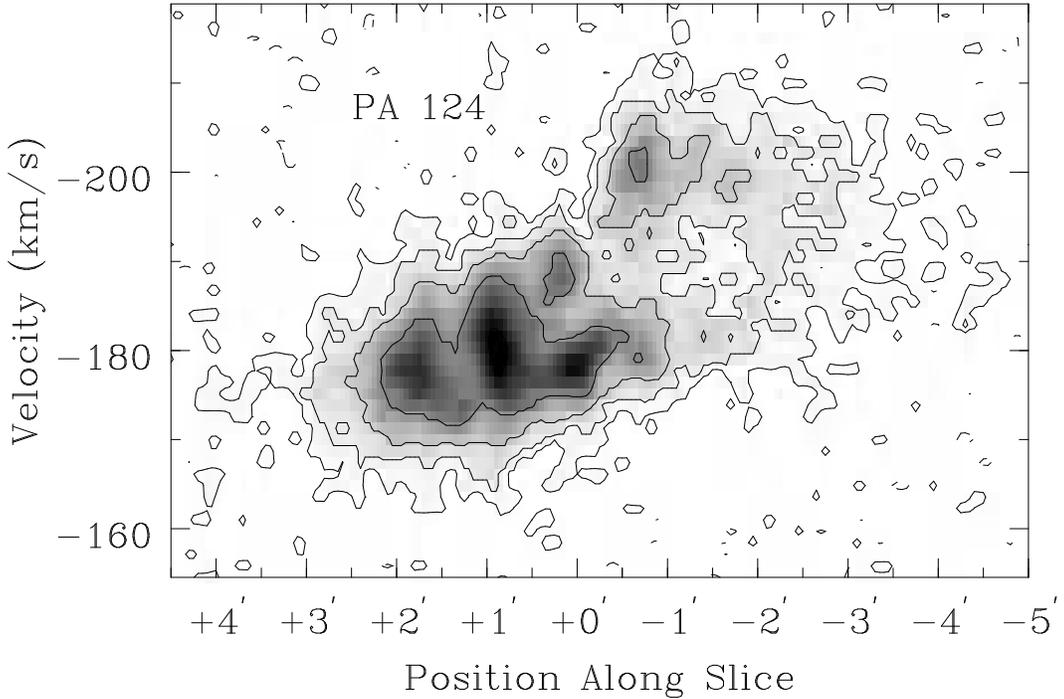}
\figcaption{Position-Velocity diagram for DDO 216, cut at a position angle of
124\arcdeg.  The contours
are 2$\sigma$, 4$\sigma$, 8$\sigma$, 16$\sigma$, and 32$\sigma$.  The ``bubble''
to the northwest corresponds
to the region where the velocity profiles are double peaked.
\label{ddo216pv}
}
\end{figure}

\begin{figure} 
 \includegraphics[scale=0.6]{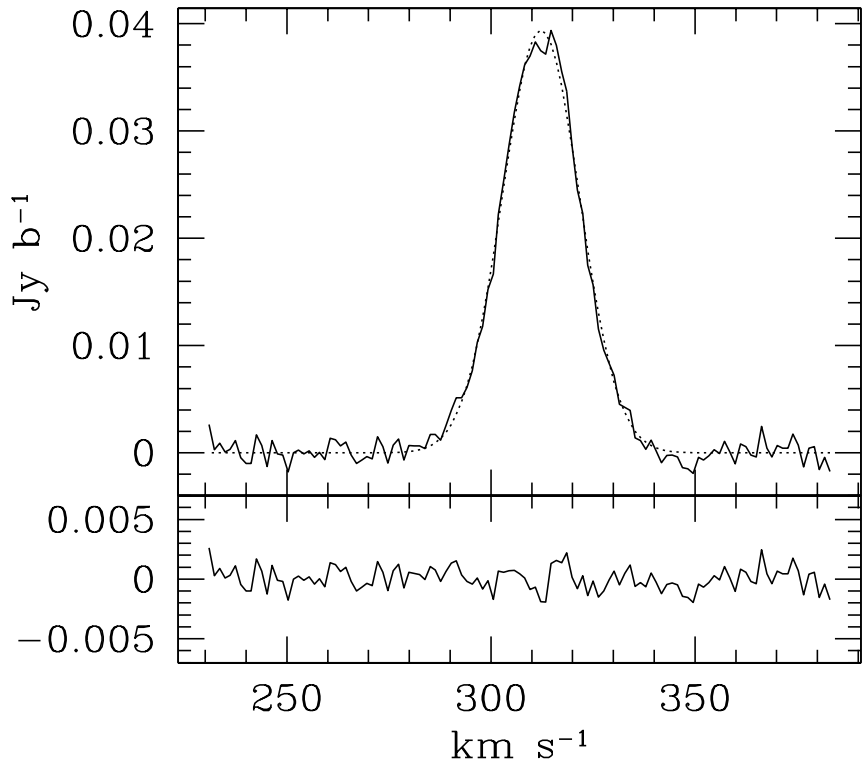}
\figcaption{Spectrum (solid line) from UGCA 292 at RA, Dec = 12:38:40.3, +32:46:08.
This spectrum is well fit by a single Gaussian component of dispersion 9.47\error
0.09 \kms (dotted line), and the residuals of the fit are plotted in the bottom
panel.  The residuals are consistent with the thermal noise level determined in
the line-free channels.
\label{puregauss}
}
\end{figure}

\begin{figure} 
 \includegraphics[scale=0.6]{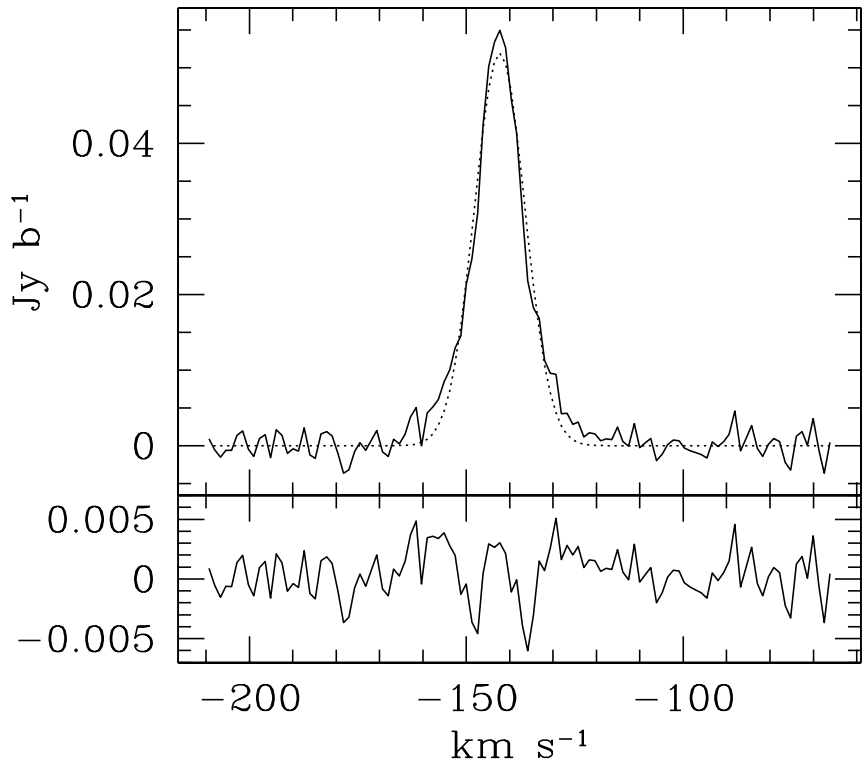}
\figcaption{Spectrum (solid line) from DDO 210 at RA, Dec = 20:46:50.2, -12:50:57.
The best fit single Gaussian model (dotted line) has a dispersion of 5.8\error
0.1 \kms, and the residuals are plotted in the lower panel.  The intensity scale
of the residuals is the same as for Figures \ref{puregauss} and \ref{h3spec}.
The f-test on the residuals indicates that, at greater than 99.9\% confidence, this
spectrum is better fit by a Gauss-Hermite polynomial with $h_3$ = 0.00\error 0.01
and $h_4$ = 0.10\error 0.01. The spectrum is also well fit by two Gaussian components with the same center
velocities but with (amplitude, dispersion) = (21\error 3 \mjb, 9.0\error 0.6
\kms) and (34\error 3 \mjb, 3.8\error 0.3 \kms).
\label{h4spec}
}
\end{figure}

\begin{figure} 
 \includegraphics[scale=0.6]{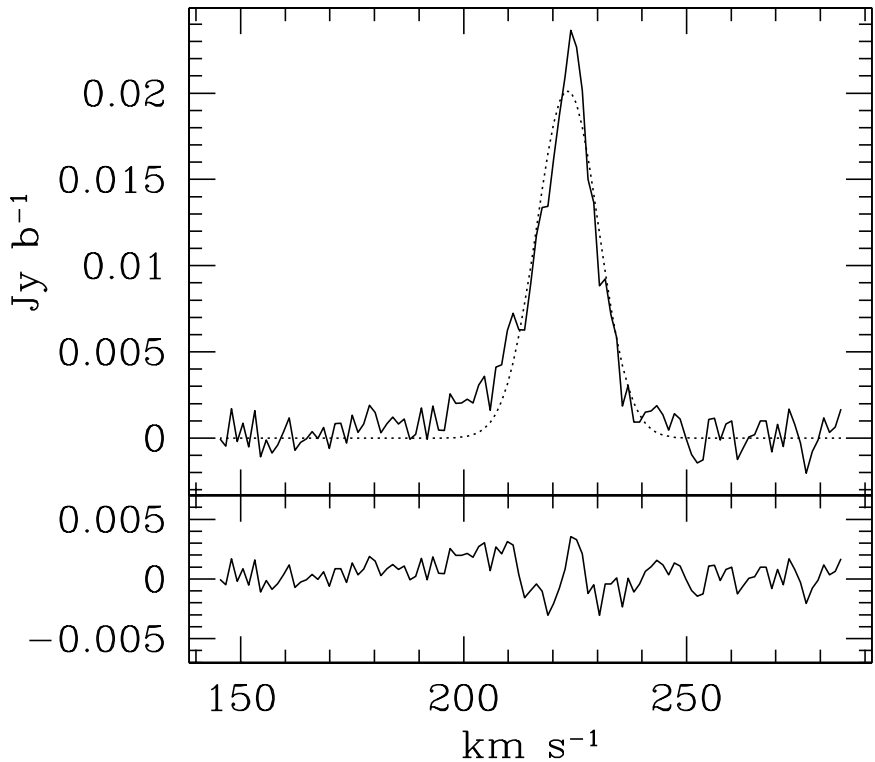}
\figcaption{Spectrum (solid line) from GR8 at RA, Dec = 12:58:39.5, +14:13:38.
The best fit single Gaussian model (dotted line) has a dispersion of 7.0\error
0.2 \kms, and the residuals are plotted in the lower panel.
The f-test on the residuals indicates that, at greater than 97.5\% confidence, this
spectrum is better fit by a Gauss-Hermite polynomial with $h_3 = -0.10$ \error 0.02
and $h_4$ = 0.12\error 0.02. The spectrum is also well fit by two Gaussian
components with (amplitude, center, dispersion) = (8\error 1 \mjb, 220\error 1
\kms, 11.9\error 0.8 \kms) and (15\error 1 \mjb, 224.3\error 0.2 \kms, 4.4\error 0.3 \kms).
\label{h3spec}
}
\end{figure}

\begin{figure} 
\figcaption{GR8, $h_3$ (boxes) on total HI column density (contours) and \halpha\
(greyscale). 
Boxes indicate the value of $h_3$ for profiles which, by the f-test, do require
the extra terms at 90\% confidence level or greater, and for which the fitted value of $h_3$ is greater than three
times its own uncertainty.  In this and subsequent box overlay figures, 
boxes are plotted
for approximately every fourth fitted spectrum (every
other one in the horizontal and vertical directions, before regridding to match
the optical pixel size).  Regions where no boxes are overlaid have a measurement of
$h_3$ consistent with zero or have profiles too weak to get a good
measurement of $h_3$.
The size of the open boxes indicates the magnitude of $h_3$ from -0.065 (smallest
boxes) to -0.245 (largest boxes), and one filled box has $h_3 = +0.08$.
The position of the spectrum in Figure \ref{h3spec} is indicated with
a cross.  The resolution of the HI data is indicated in the upper left corner.
\label{gr8h3hiha}
}
\end{figure}

\begin{figure} 
\figcaption{GR8, $h_4$ (boxes) on HI total column density (contours) and \halpha
(grayscale).  
As for Figure \ref{gr8h3hiha}; the smallest boxes have $h_4$ = 0.10 and the largest
boxes have $h_4$ = 0.16.  Approximately one fourth of the fitted spectra are plotted.
\label{gr8h4hiha}
}
\end{figure}

\clearpage

\begin{figure} 
\figcaption{UGCA 292, $h_3$ (boxes) on HI total column density (contours) and \halpha
(grayscale, from van Zee 2000). 
The position of the spectrum in Figure \ref{puregauss} is indicated with a cross.
The open boxes correspond to $h_3$ values from
$-0.039$ (smallest boxes) to $-0.055$ (largest boxes), and the filled boxes
correspond to $h_3$ values from 0.040 (smallest boxes) to 0.081 (largest boxes).
\label{cvnah3hiha}
}
\end{figure}

\begin{figure} 
\figcaption{UGCA 292, $h_4$ (boxes) on HI total column density (contours) and \halpha
(grayscale).  
The boxes correspond to $h_4$ values from
0.03 (smallest boxes) to 0.10 (largest boxes), and other items are as for Figure \ref{cvnah3hiha}.
\label{cvnah4hiha}
}
\end{figure}

\begin{figure} 
\figcaption{DDO 210, $h_3$ (boxes) on HI total column density (contours) and
\halpha (greyscale).
The position of the spectrum in Figure \ref{h4spec} is indicated with a cross.
The open boxes correspond to $h_3$ values from
$-0.06$ (smallest boxes) to $-0.1$ (largest boxes), and the filled boxes
correspond to $h_3$ values from 0.05 (smallest boxes) to 0.12 (largest boxes).
\label{ddo210h3hiha}
}
\end{figure}

\begin{figure} 
\figcaption{DDO 210, $h_4$ (boxes) on HI total column density (contours) and
\halpha (greyscale).
The boxes correspond to $h_4$ values from
0.05 (smallest boxes) to 0.15 (largest boxes), and other items are as for Figure
\ref{ddo210h3hiha}.
\label{ddo210h4hiha}
}
\end{figure}

\clearpage

\begin{figure} 
 \includegraphics[scale=1.0,bb=117 350 400 638]{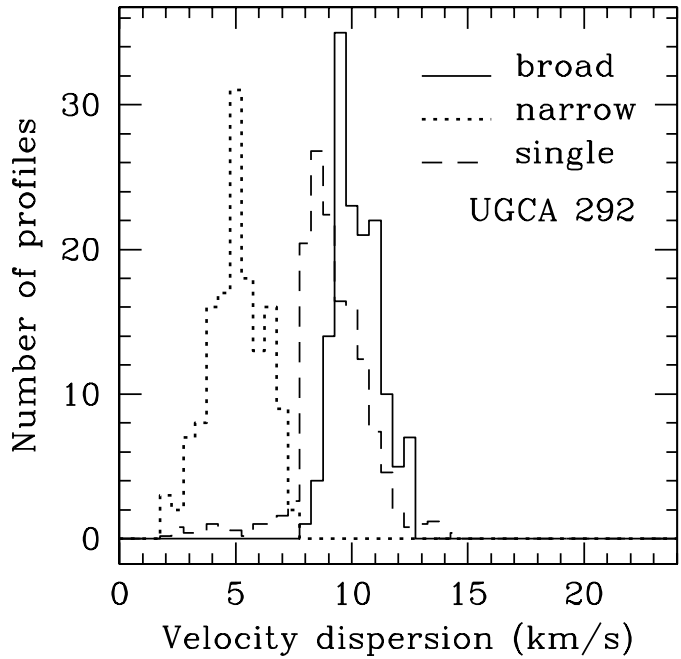}
\figcaption{Fitted dispersions in UGCA 292.  For profiles that require two components at
the 90\% or higher confidence level of the f-test, the dotted line shows the
dispersions of the narrower component and the solid line corresponds to the
dispersions of the broader component.  The dashed line shows the dispersions of
profiles which are adequately described by a single Gaussian component, and the
vertical scale is reduced by a factor of 5.0 for this set.
\label{cvnahist}
}
\end{figure}

\begin{figure} 
 \includegraphics[scale=1.0,bb=117 350 400 638]{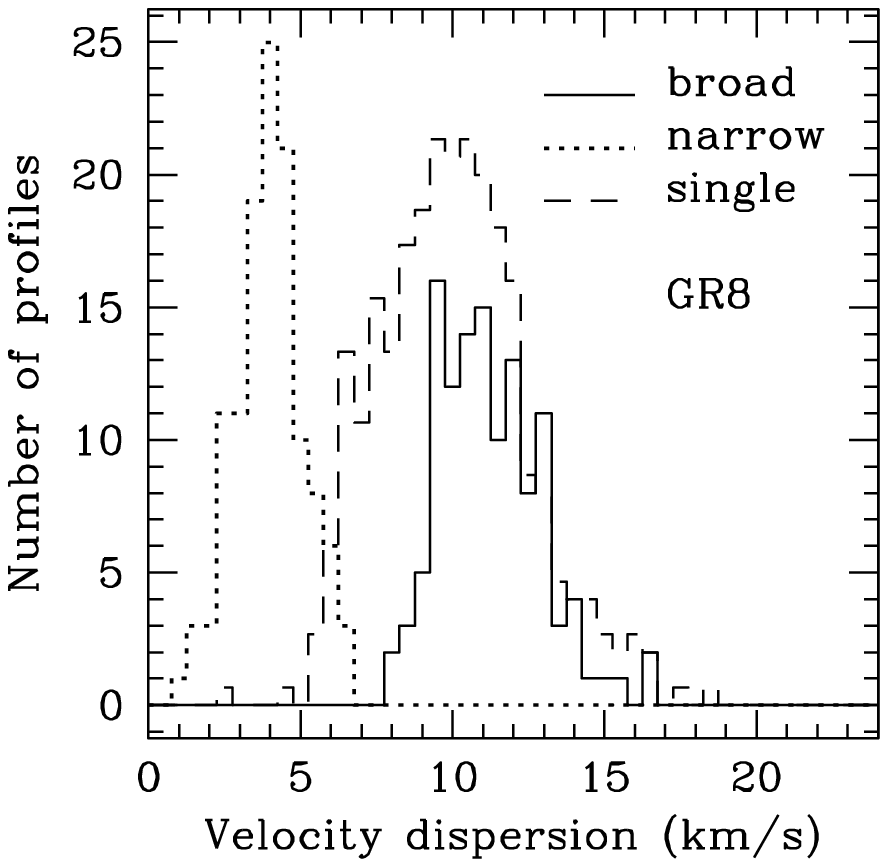}
\figcaption{Fitted dispersions in GR8.  Similar to Figure \ref{cvnahist}, but the
vertical scale of the single-component profiles is reduced by a factor of 1.5.
\label{gr8hist}
}
\end{figure}

\begin{figure} 
 \includegraphics[scale=1.0,bb=117 350 400 638]{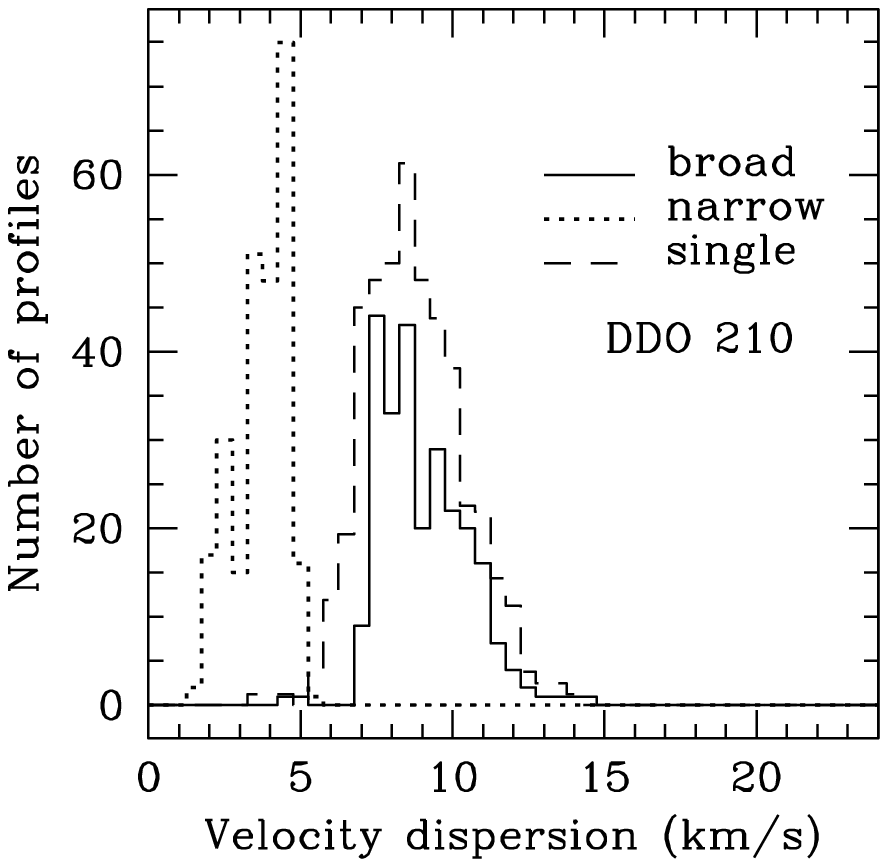}
\figcaption{Fitted dispersions in DDO 210.  Similar to Figure \ref{cvnahist}, but the
vertical scale of the single-component profiles is reduced by a factor of 1.6.
\label{ddo210hist}
}
\end{figure}

\begin{figure} 
\figcaption{GR8, column density of the narrower Gaussian component (boxes) on total
HI column density (contours) and \halpha.   Only profiles which require the two
components at greater than 90\% confidence are indicated.  The smallest box corresponds to a
column density of approximately 5\e{19} \persqcm\ and the largest to 7\e{20}
\persqcm.  Uncertainties in these column densities are
estimated around 10--15\% for the profiles with peak signal-to-noise ratios of
30, increasing to 30--40\% at signal-to-noise ratio of 11 (Table
\ref{monte-gauss}).  The position of the spectrum of Figure \ref{h3spec} is indicated by a
cross, and the resolution of the HI data is shown in the upper left corner.
As for Figures \ref{gr8h3hiha} through \ref{ddo210h4hiha}, \ref{cvnanarrow} and
\ref{ddo210narrow}, every fourth profile
(before regridding) is plotted.
\label{gr8narrow}
}
\end{figure}

\begin{figure} 
\figcaption{UGCA 292, column density of the narrower Gaussian component (boxes) on total
HI column density (contours) and \halpha.   The smallest box corresponds to a
column density of approximately 9\e{19} \persqcm and the largest 1.5\e{21}
\persqcm.   Uncertainties in the column densities are as for Figure
\ref{gr8narrow}.
The position of the spectrum of Figure \ref{puregauss} is indicated by a
cross, and the resolution of the HI data is shown in the upper left corner.
\label{cvnanarrow}
}
\end{figure}

\begin{figure} 
\figcaption{DDO 210, column density of the narrower Gaussian 
component (boxes) on total HI column density (contours) and \halpha (greyscale).
The smallest box corresponds to approximately 3\e{19} \persqcm and 
the largest box 1.0\e{21} \persqcm.  Uncertainties in these values are
as for Figure \ref{gr8narrow}.  The position of the spectrum of Figure \ref{h4spec} is indicated by a
cross, and the resolution of the HI data is shown in the upper left corner.
\label{ddo210narrow}
}
\end{figure}

\begin{figure} 
 \includegraphics[scale=0.5]{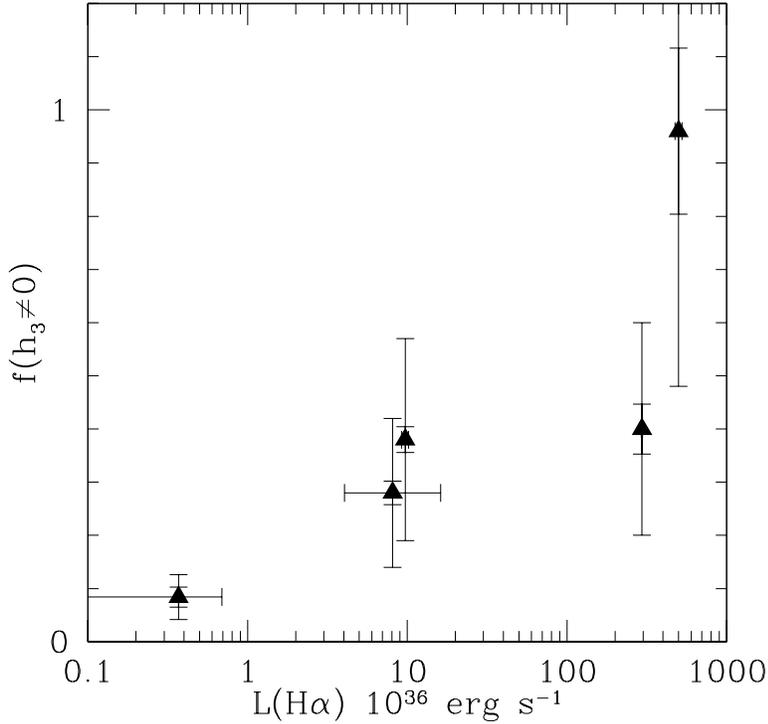}
\figcaption{\halpha luminosity vs. the fraction of asymmetric profiles \fhthree\ for the five
galaxies in Table \ref{h3vshatable}.
Two error estimates are shown for \fhthree: the smaller set of error bars
corresponds to the assumption of Poisson statistics as shown in the Table.
These values are likely to underestimate the true uncertainties because adjacent
pixels do not contain completely independent data (though the fits are made
independently).
The larger set of error bars is a somewhat arbitrary assumption of a 50\% error in \fhthree.
\label{h3vssfr}
}
\end{figure}

\begin{figure} 
 \includegraphics[scale=0.5]{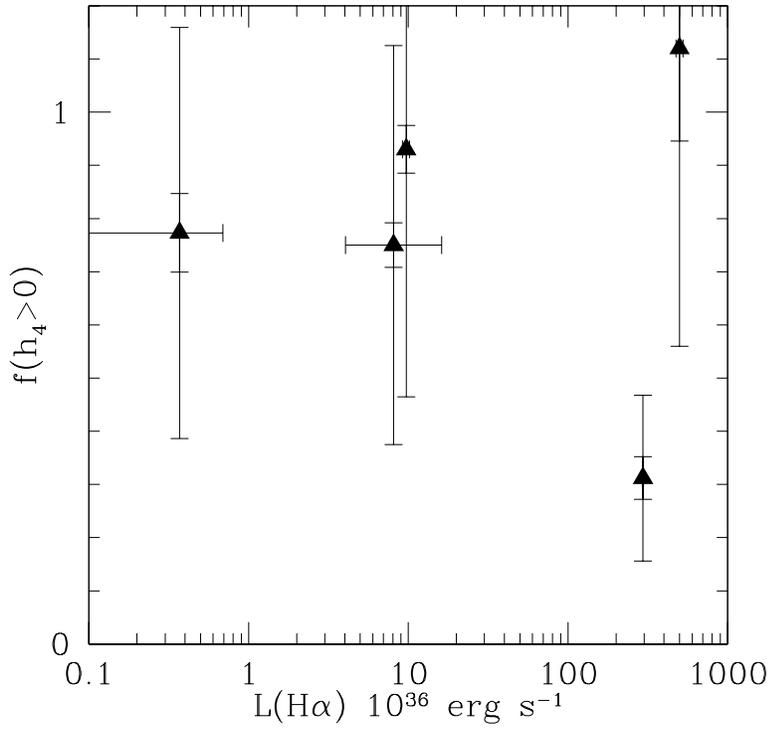}
\figcaption{Similar to Figure \ref{h3vssfr}, but for $h_4 > 0$.
\label{h4vssfr}
}
\end{figure}

\end{document}